\documentclass[twoside,reqno]{article}
\usepackage{epsfig,cite,colordvi}
\usepackage{graphicx}
\usepackage{subfig}
\usepackage{url}
\usepackage{amssymb,amsmath,amscd,epsf}
\usepackage{times}
\usepackage{mathptmx}

\def\etal{\textit{et al.}}
\def\ie{\textit{i.e.}}
\def\eg{\textit{e.g.}}

\begin{document}

\title{Challenges for Modeling Nuclear Structure:\\ 
\large Are the Proton and Neutron Masses and A-body Interactions Relevant?}

\author{V. G. Gueorguiev$^1$, P. Navr\'atil$^2$, J. P. Vary$^3$, J. P. Draayer$^4$, F. Pan$^5$\\ \\
\it \small $^1$Institute of Nuclear Research and Nuclear Energy,\\
\it \small Bulgarian Academy of Sciences, Sofia 1784, Bulgaria\\
\it \small $^2$Lawrence Livermore National Laboratory, Livermore, California, USA\\
\it \small $^3$Iowa State University, Ames, Iowa, USA\\
\it \small $^4$Department of Physics and Astronomy, Louisiana State University,\\ 
\it \small Baton Rouge, Louisiana, USA\\
\it \small $^5$Department of Physics, Liaoning Normal University, \\
\it \small Dalian 116029, P.R. China}
\date{\small Contribution to the XXIX International Workshop on Nuclear Theory,\\
\small June 20 - 26, 2010, Rila Mountains, Bulgaria}
\stepcounter{footnote}
\footnotetext{On leave of absence from the Institute for Nuclear Research and Nuclear Energy, Bulgarian Academy of Science, Sofia, Bulgaria. For current mailing address look up the APS members directory or send e-mail to vesselin at mailaps.org.}
\maketitle

\begin{abstract}
We discuss some of the challenges that future nuclear modeling may face in order to improve the description of the nuclear structure. One challenge is related to the need for A-body nuclear interactions justified by various contemporary nuclear physics studies. Another challenge is related to the discrepancy in the NNN contact interaction parameters for $^3$He and $^3$H that suggests the need for accurate proton and neutron masses in the future precision calculations.
\end{abstract}

\section{Introduction}

The high precision, QCD derived, nucleon interaction that describes the NN-scatering phase shifts, the deuteron, and the light s- and p-shell nuclei points to the necessity of NNN-interaction terms \cite{Machleidt, Petr}. Thus the conventional two-body interaction paradigm is challenged and the need of 3-body and possibly A-body interaction define a new research frontier. The structure of the three-body terms has been studied previously using the meson exchange theory \cite{TM79}. However, with the advance of the Chiral Perturbation Theory (ChPT) \cite{TM'99,TM'01} the structure of the three-body terms is better justified using QCD. While studying the parameters of the 3-body contact terms \cite{Petr} one faces a discrepancy in the NNN contact interaction parameters needed to fit $^3$He and $^3$H that could be viewed as an argument towards implementing the accurate proton and neutron masses in future precision calculations. 

Higher many-body interaction terms (e.g. NNNN-interaction terms) are also part of the interaction as derived from QCD via ChPT \cite{Epelbaum}. The Okubo-Lee-Suzuki (OLS) effective interaction method, employed in solving the nuclear many-body theory, also introduces interaction terms beyond the common 2-body interaction \cite{Okubo, Lee-Suzuki}. All this seems to be pointing to the need of A-body interactions for the description of the nuclear structure. It also raises the question about the importance of the A-body interactions in very heavy nuclei. Fortunately, there is an exactly solvable A-body model - \textit{the extended pairing model} - that is applicable as an A-body interaction to very heavy nuclei; therefore, it can help to address this question 
\cite{Feng,VGG2004,EPJ05}.

In the next section we briefly discuss the microscopic nuclear physics hamiltonian; the types of the high-precision NN-interaction potentials and their failure to properly account for the structure  of the nuclei with more than two nucleons. In Sec. \ref{section2} we discuss the discrepancy in the values of the $c_D$ and $c_E$ NNN-intercation parameters \cite{Petr} and try to argue that the application of high-presision nucleon potentials needs more appropriate nucleon masses for high-precision description of the A=3 systems. In Sec. \ref{section3} we further extend our argument for A-body nuclear interactions by using the modern OLS effective interaction in finite model space method. In Sec. \ref{section4} we briefly discuss the well-know 2-body pairing interaction and its exact solution as a prelude to the A-body Extended Pairing Interaction (EPI); then we discuss the results of apply the EPI to  few long isotope chains like Sn, Yb, and Pb nuclei. Last section is our conclusion about the needs of the future nuclear structure modeling.  

\section{Modeling the Nuclear Interactions}
\label{section1}

Unlike the electromagnetic and the gravitational intersection, the mathematical form of the nuclear interaction has been very elusive. It is now clearly understood that this is due to the fact that the nuclear interaction arises nontrivially from the quark structure of the nucleons and thus related to the theory of the QCD. However, the absence of  a closed form interaction has not hindered researchers from modeling the structure of nuclei. The field has advanced significantly, based on general quantum mechanical principals and techniques. In particular, the microscopic approach has been very successful especially with the advance of computational techniques and computer power that have allowed for the construction of effective high-precision meson and/or QCD derived NN-potentials. The free parameters of the high-presision NN-potentials are usually fixed by the experimental two-nuceon scattering data and describe the 2-body system extremely well. Unfortunately, these potentials produce unsatisfactory description of the 3- and 4-body systems.

\subsection{The Nuclear Shell-Model Hamiltonian}

A nuclear many-body system near equilibrium can be viewed as subject to a mean field Harmonic Oscillator (HO) potential:
$ H_{0}=\frac{\vec{p}^2}{2m}+\frac{1}{2}k^2\vec{x}^2.$
It is well know that one can understand the magic numbers and the shell structure of nuclei within the 3-dimensional HO approximation plus a spin-orbit potential \cite{Haxel&Jensen}. Using the HO single-particle states  one can write a general  Hamiltonian with one- and two-body terms:
\begin{equation}
H=\sum_{i}\varepsilon_{i}a^{+}_{i}a_{i}+
\frac{1}{4}\sum_{i,j,k,l}V_{ij,kl}a^{+}_{i}a^{+}_{j}a_{k}a_{l}.
\label{H2body}
\end{equation}
Here, $a_{i}$ and $a^{+}_{j}$ are fermion annihilation and creation operators, $\varepsilon_{i}$ single-particle energies, and $V_{ij,kl}=\left< ij|V|kl\right>$ two-body interaction matrix elements and the index $i$ labels the single particle levels. Despite the significant symmetry relations, \eg  $~\varepsilon_{jm}=\varepsilon_{jm'}$ due to rotational symmetry and $V_{ij,kl}=V_{kl,ij}=-V_{ji,kl}=-V_{ij,lk}$ due to the fermion exchange properties and the hermition requirement on the energy operator,  the number of independent parameters is often more than a dozen - usually it is of order of few hundred  for the valence NN interactions alone. The independent parameters of the interaction (\ref{H2body}) are often fitted to experimental data by starting with some initial values that come from a relevant theory or model. 

\subsection{Problems with the High-Precision NN-Potentials}

Many of the high-precision NN-potentials commonly used to build the microscopic interactions for multi-nucleon systems have very complicated but methodically developed structure in terms of spin, iso-spin, and angular momentum components although sometimes there is a very complicated radial dependence. For example, the Argonne V18 potential has 18 different terms \cite{AV18}. Other potentials use non-local terms e. g. CD-Bonn \cite{CD-Bonn} and Nijmegen \cite{Nijmegen}. However, when applied to A$>$2 systems all of these potentials have a serious difficulties that were eventually overcome by using three-body interactions \cite{TM'99,UIX, Wiringa}.

By the end of the twentieth century it become clear that a two-body interaction by itself is inadequate even for the description of the lightest nuclei $2<A<5$. Comparative studies of various potentials, such as AV18, Nijmegen, CD-Bonn, and N$^3$LO, with or without three body terms have demonstrated the inadequacy of the pure two-body interactions and the need for three-body interaction terms \cite{Wiringa, Entem&Machleidt}. For example, all these interactions (AV18, Nijmegen, CD-Bonn, and N$^3$LO) describe very well the deuteron properties such as binding energy, radius, and quadruple moment but fail by more than 0.5 MeV to reproduce the binding energy of triton 
\cite{Entem&Machleidt} and underbind $^4$He by more than 4 MeV \cite{Wiringa}.

Although the meson-exchange approach was successful, it was clear that this phenomenological models should be derived from the underling QCD. Thus the ChPT approach became a prominent technique that produced the high-precision NN-potential  N$^3$LO and then guided the researchers into the structure of the NNN- and NNNN-interactions 
\cite{Entem&Machleidt,NNN_N2LO,Epelbaum}.

\section{Light Nuclei and the Parameters of the NNN-body Interaction}
\label{section2}

The use of the ChPT in the derivation of the nucleon interactions from QCD assisted in the determination of the mathematical form of various interaction terms along with the relevant parameters. Unfortunately, parameters related to contact terms in the interaction could not be determined. Thus the $c_D$ and the $c_E$ strengths of the two-nucleon contact interaction with one-pion exchange to a third nucleon and the three-nucleon contact interaction are identified freedoms at the present time in the effective ChPT interaction. As such they need to be fixed by comparison with experiment.

\subsection{Binding Energy of $^3$H, $^3$He, and $^4$He}

\begin{figure}[htb]
\centering
\includegraphics[scale=0.3]{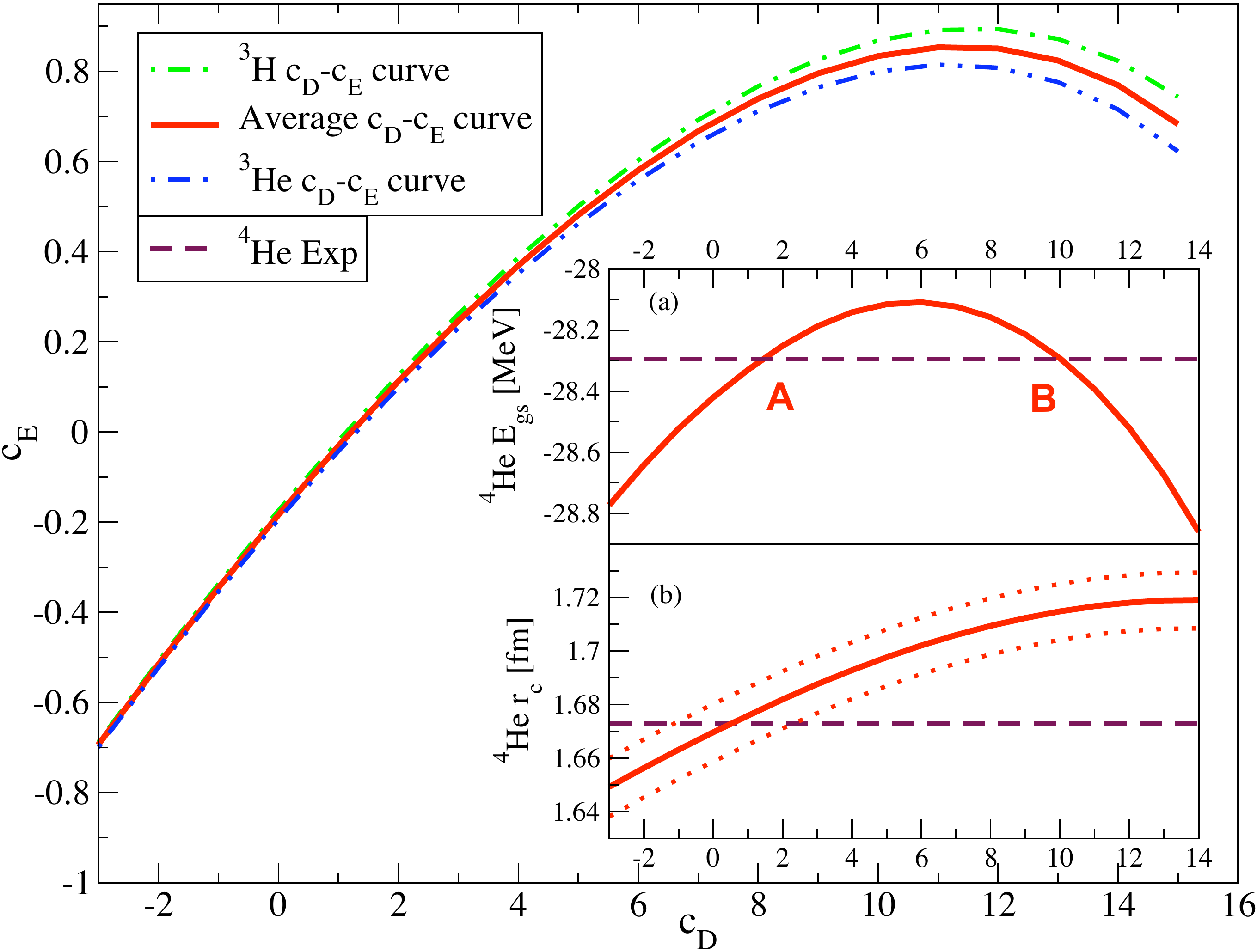}
\caption[]{Relations between $c_D$ and $c_E$ for which  the 
binding energy of $^3$H ($8.482$ MeV) and  $^3$He ($7.718$ MeV) are reproduced. 
(a) $^4$He ground-state energy along the averaged $c_D-c_E$ curve. 
The experimental $^4$He binding energy ($28.296$ MeV) is reproduced to within ~0.5 MeV over the entire range depicted.
(b) $^4$He charge radius $r_c$ along the averaged $c_D-c_E$ curve. Dotted lines represent 
the $r_c$ uncertainty due to the uncertainties in the proton charge radius.}
\label{CDCEmtAB}
\end{figure}

In order to determine the $c_D$ and $c_E$ parameters of the interaction one searches for the parameter values that reproduce the binding energy of $^3$H and $^3$He within 0.5 keV of the experimental values \cite{Petr}. As seen from Figure \ref{CDCEmtAB} there are two $c_D-c_E$ curves that unfortunately do not intersect. In order to further narrow down the range of  $c_D$ values one considers the averaged $c_D-c_E$ curve and evaluates the binding energy of the $^4$He system which results in two possible physical regions denoted by A and B on inset (a) of Figure \ref{CDCEmtAB}. Finally, the charge radius of $^4$He points to the region A as the reasonable range of values for the $c_D$ parameter while the $c_E$ parameter is determined by the averaged $c_D-c_E$ curve.

Conceptually, there are three important concerns: First, the ChPT NN-potential was one order higher than the NNN-potential and no NNNN-potential was included. That is, the high-precision NN-potential was N$^3$LO (next-to-next-to-next-to-leading order) \cite{Entem&Machleidt} while the ChPT NNN-potentail was at the N$^2$LO order \cite{NNN_N2LO} and the NNNN-potential \cite{Epelbaum} was not yet readily available. The second concern is that the range of the 3-body interaction parameter $c_D$ is determined by the properties of the 4-body system $^4$He; this, however, was resolved by a later study that used the $\beta^-$ decay of $^3$H into  $^3$He  and confirmed the physically relevant region A for the parameter $c_D$  \cite{3Hto3He}. The third concern  is related to the fact that these are high-presision studies and at this level of accuracy the difference between the proton and nucleon mass could be important for the A=3 systems. 

\subsection{$^3$H and $^3$He Systems with Modified Nucleon Mass}

It is clear from Figure \ref{CDCEmtAB} that the $c_D-c_E$ curves for $^3$H and $^3$He do not intersect in the physically relevant region ($-1 < c_D < 1$). This could be attributed to the absence of the T=3/2 channel in these first calculations. The slight difference in the $c_D$ value as suggested by $^4$He binding energy and its charge radius could be attributed to the  inconsistency of the different interaction terms, \ie  ~ NN-terms are at N$^3$LO level while the NNN-terms are at N$^2$LO level and the NNNN-terms are not present at all. Another source of these discrepancies could be the conventional use of equal masses for protons and neutrons. We will discuss this option in more detail below. 

\begin{figure}[htb]
\centering
\includegraphics[scale=0.4]{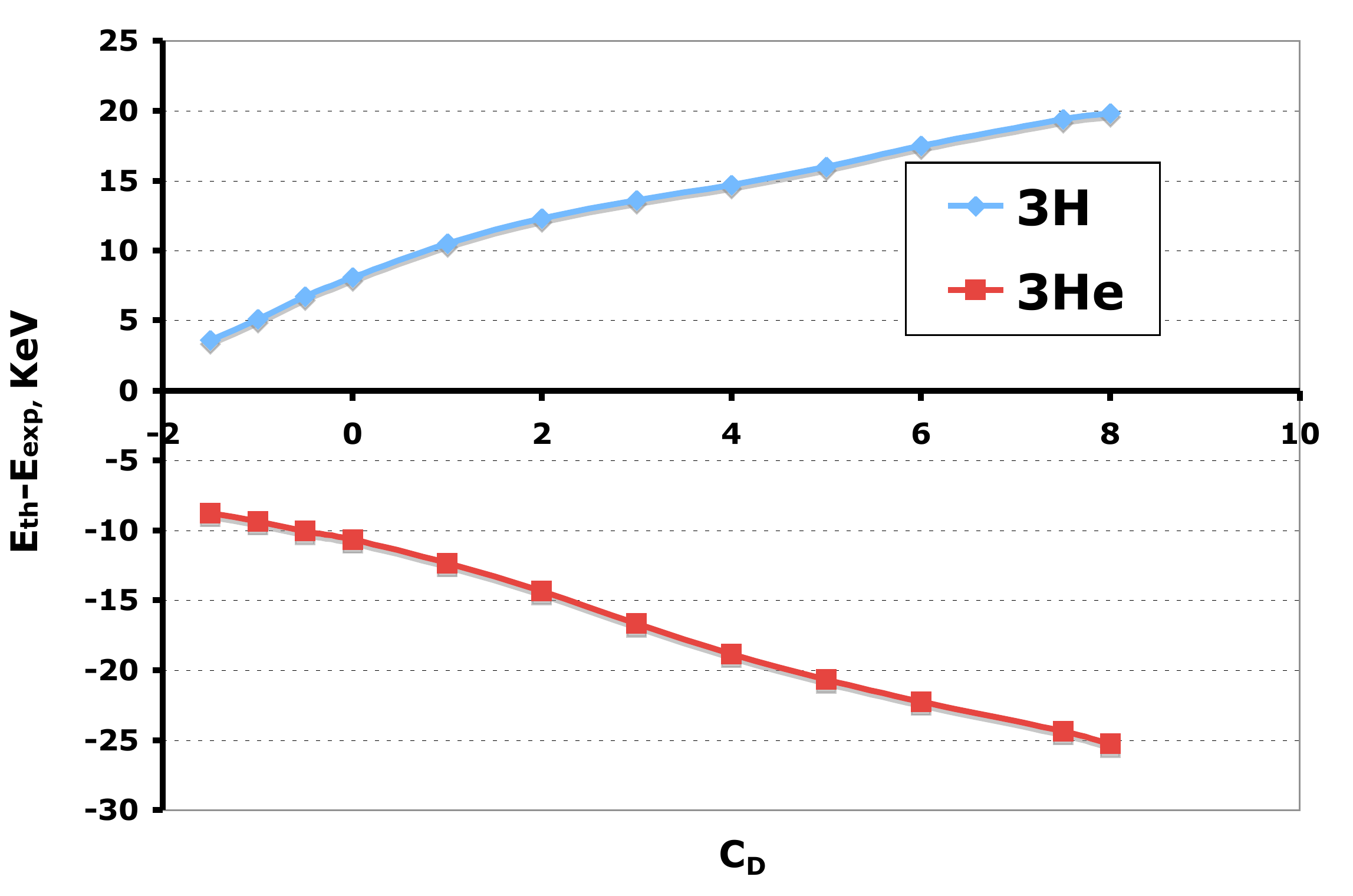}
\caption[]{Deviation of the binding energy of the three-nucleon systems as computed along the averaged $c_D-c_E$ curve.}
\label{dBE_CD}
\end{figure}

As seen from Figure \ref{dBE_CD} the binding energy deviation range is $15 < \delta K< 25 $ keV within the physically relevant region ($-1 < c_D < 1$). This is within the accuracy of the kinetic energy K as evaluated for equal mass nucleons $m=m_n=m_p=(m_n+m_p)/2$. Since the averaged relative kinetic energy for the three-nucleon system is about $K\approx 37 $ MeV and the relative nucleon mass deviation $\delta m /m$ is $\approx 0.7 \times 10^{-3}$ with $\delta m =(m_n-m_p)/2$, we have:
\begin{equation}
K=\frac{m}{2}v^2  \Rightarrow ~
\delta K=K \frac{\delta m}{m} \approx 26~\textrm{keV}
\label{dK}
\end{equation}
This shows that there is not a single $c_E$ value that will result in perfect description of the $^3$H and $^3$He systems. One could hope that including the T=3/2 channel would improve the situation. Alternatively, with this level of precision, we are led to investigate corrections to the conventional  $m_n = m_p$ approximation. One can test the sensitivity to the conventionally used nucleon mass by changing it to a more appropriate value 
\cite{Kamuntavicius'99}. 
\begin{equation}
m=\frac{1}{A}(Z m_p+(A-Z) m_n).
\label{Nmass_new}
\end{equation}

\begin{figure}[htb]
\centering
\includegraphics[scale=0.4]{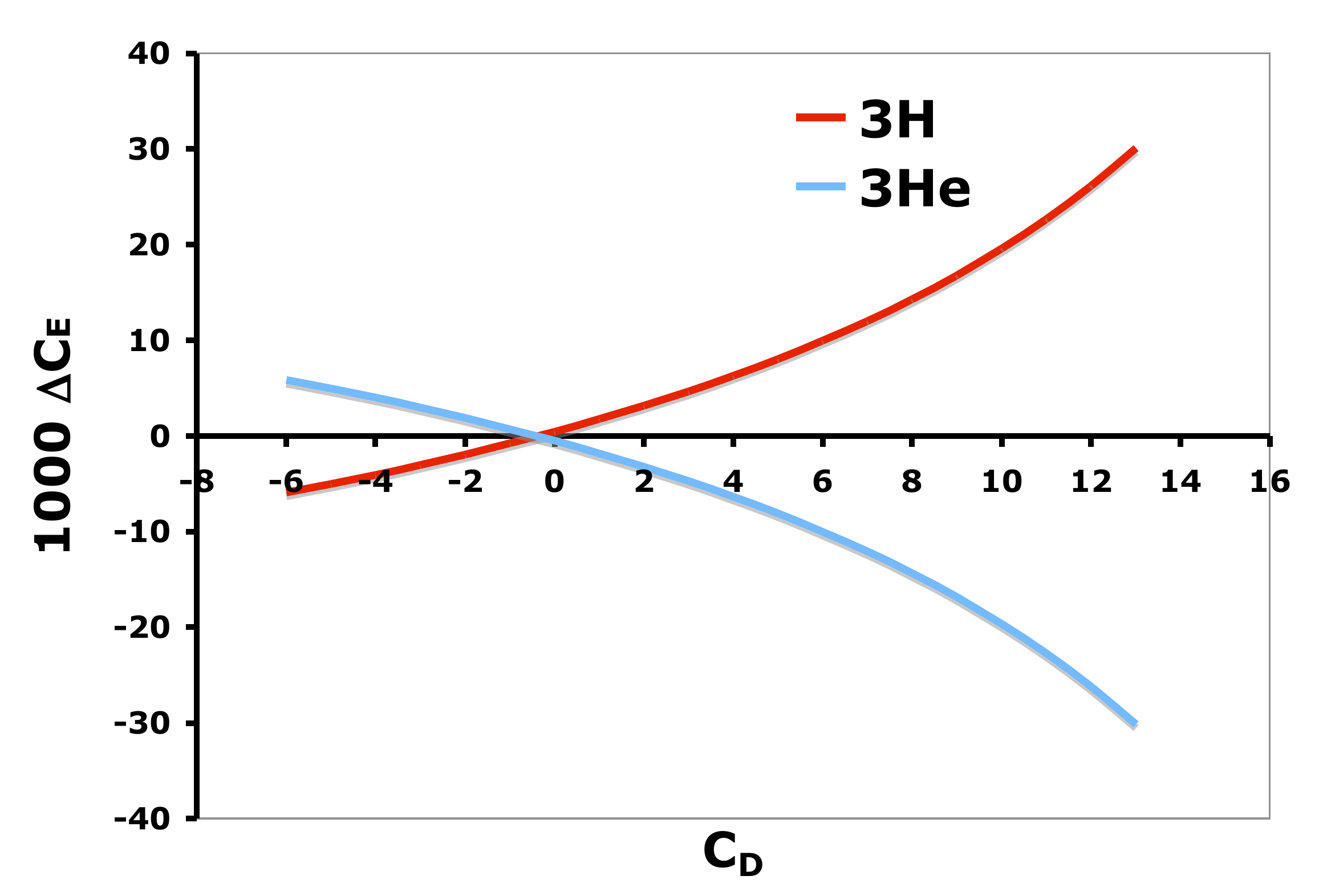}
\caption[]{Intersecting $c_D-c_E$ curves where the binding energy of $^3$H ($8.482$ MeV) and  $^3$He ($7.718$ MeV) are reproduced when using modified nucleon mass as suggested by  (\ref{Nmass_new}).}
\label{dCE_CD_anu}
\end{figure}

If one repeats the calculations related to Figure \ref{dBE_CD} but by employing the nucleon mass value as suggested by (\ref{Nmass_new}), one obtains an interesting result shown in Figure \ref{dCE_CD_anu}.
From Figure \ref{dCE_CD_anu} is clear that there are unique $c_D$ and $c_E$ values where both binding energies can be reproduced exactly. The  $c_D$ is in agreement with the $c_D$ value estimated from the charge radius of $^4$He (see Figure \ref{CDCEmtAB}(b)). Since, $^4$He has same number of protons and neutrons, there is no nucleon mass adjustment for this system, so results shown for $^4$He on Figure \ref{CDCEmtAB} are still valid. Perhaps  by incorporating NNNN-interaction \cite{Epelbaum} the binding energy for the $^4$He would agree better with the $c_D$ value suggested by the three-nucleon system as calculated with a modified nucleon mass and the $\beta^-$ decay of $^3$H \cite{3Hto3He}.   

\section{Beyond the 2-body Interaction - Effective Interactions in a Finite Model Space}
\label{section3}

In the previous section we discussed results obtained by using QCD derived interactions and the role of the NNN-interaction in the description of the light nuclei. Clearly 3- and 4-body interaction terms are predictions of the ChPT. Thus A-body interactions can be viewed as real physical interactions within the ChPT approach to nuclei. However, there is another way to arrive at A-body interactions which are phenomenological effective interactions since they are related to our inability to handle  interacting systems in infinite Hilbert spaces \cite{AbodyHeff}. Since the quality of a model is judged by its ability to reproduce the experimental data, as far as computational models are concerned, an A-body interaction which gives results that agree well with the data is physically relevant as well.

In practice, we are computationally limited to a finite subspace of the infinite Hilbert space of the full quantum many-body problem. The subspace that we can access is defined by finite set of convenient many-body basis states. For a suitable choice of basis we hope to have good overlaps with low-lying physical states of the system under study. If we imagine the exact solutions are available for analysis and apply a unitary transformation to those eigenstates, we can produce a transformed set of solutions maximally overlaping with our chosen basis space.

\begin{figure}[htb]
\centering
\includegraphics[scale=0.4]{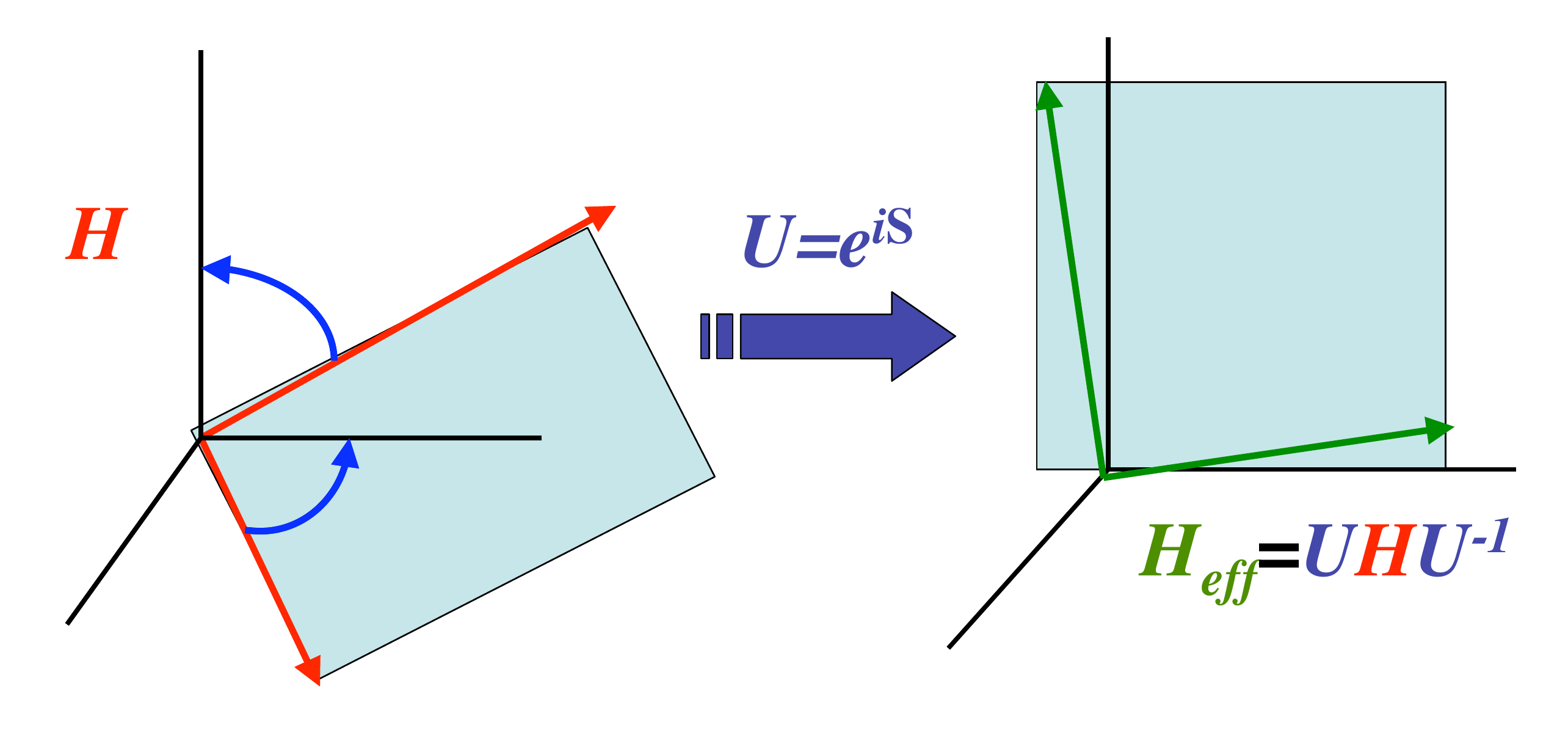}
\caption[]{Geometrical interpretation of the Okubo-Lee-Suzuki transformation method for construction of effective Hamiltonian operators.}
\label{OLS_H_eff}
\end{figure}

For example, one may be interested in the lowest two energy states of a system, as shown in Figure \ref{OLS_H_eff} left, but would like to have some unitarily transformed version of these states that have maximal overlap with the two basis states that define the plane of the page (Figure \ref{OLS_H_eff} right). By finding the relevant unitary transformation U, one can define an effective Hamiltonian that would have the lowest two states as desired.  Then this effective Hamiltonian could be used in the calculations of more complicated multi-particle systems, \ie ~ one would find the unitarily transformed Hamiltonian that describes very well the low-energy states of a 2-body system in a mean field but within a Fock space that would be used later for an A-body system. Unfortunately, this transformation will turn any one- and two-body potential into a 
many-body effective interaction: 
\begin{eqnarray}
\frac{1}{2}\sum_{i \ne j}^{A} V_{ij}~\xrightarrow{U=e^{i S}} 
V_{eff}=\sum_{k=1}^A\frac{1}{k!}\sum_{i_{1}, \cdots, i_{k}}^A V_{i_{1}, \cdots, i_{k}} \nonumber
\label{H2bodyToAbody}
\end{eqnarray}
This way the two-body Hamiltonian (\ref{H2body}) becomes an 
A-body Hamiltonian:
\begin{equation}
H=\sum_{i}\varepsilon_{i}a^{+}_{i}a_{i}+
\sum_{k=2}^{A}\frac{1}{(k!)^2}
\sum_{
\begin{split}
_{i_{1},\ldots,i_{k},}\\
^{j_{1},\ldots, j_{k}} 
\end{split}}
V_{i_{1},\ldots,i_{k},j_{1},\ldots, j_{k}}a^{+}_{i_1}\cdots a^{+}_{i_k}a_{j_1}\cdots a_{j_k}.
\label{HAbody}
\end{equation}


For A$>$4, it seems impractical at present to obtain the structure of the A-body interactions as derived from ChPT as it was previously done for the NNN- and the NNNN-interaction terms. Before embarking on the extensive undertaking required for including higher-body effective interactions, it would be very helpful to investigate a simple exactly solvable A-body interaction model that has few parameters and is applicable to real A-body systems.


\section{The Extended Pairing Model}
\label{section4}

In order to study the relevance of the A-body interactions one should use the general form of the interaction and to try to determine some of the A-body interaction strengths since it seems impractical at present to be able to obtain the structure of the A-body interactions from ChPT for A$>$4. Therefore, as we reasoned earlier one needs simple exactly solvable A-body interaction with few parameters that can be adjusted to the experimental data. Fortunately, there is such an interaction - \textit{ the Extended Pairing Interaction} (EPI) \cite{Feng}. The discovery of this exactly solvable model was a result of research into the solution of the two-body proton-neutron pairing which turned to be exactly solvable as well \cite{Dukelsky}. However, for our purpose the justification, of the A-body EPI Hamiltonian, is the need for simplicity: thus, one can set all the unknown interaction strengths $V_{i_{1},\ldots,i_{k},j_{1},\ldots, j_{k}}$ in (\ref{HAbody}) to be equal to single strength $G$ and to consider only pairs of fermion particles $b^{+}_{i}=a^{+}_{i\uparrow}a^{+}_{i\downarrow}$:

\begin{equation}
H=\sum_{i}2\varepsilon_{i}n_i-G\sum_{i,j}b^{+}_{i}b_{j}-
\sum_{k=2}^{A}\frac{G}{(k!)^2}
\sum_{{i_{1}, \cdots, i_{2k}}}
b^{+}_{i_1}\cdots b^{+}_{i_k}b_{i_{k+1}}\cdots b_{i_{2k}}.
\label{H_extPM}
\end{equation}
Here $n_i$ counts the number of pairs on the $i$-th level; thus, the value 2 in front of the single particle energy $\varepsilon_{i}$.  If one considers a system of only one pair of particles then the $k>1$ terms in (\ref{H_extPM}) disappear since their matrix elements are zero in the one-pair basis. Thus, one gets the standard pairing Hamiltonian:
\begin{equation}
H_{P}=\sum_{j}2\varepsilon_{j}n_j-g\sum_{jj'}A^{+}_{j}A_{j'}, ~~
A^{+}_{j}=\sum_{m>0}a^{+}_{j m}a^{+}_{j -m}
\label{H_P}
\end{equation}

This 2-body Hamiltonian, however, is exactly solvable even for systems with more than one pair since it can be viewed as Richardson-Gaudin model \cite{RG-models}. For example, the relevant equations for the proton-neutron $T=1$ pairing that were given in Ref. \cite{Dukelsky} as well as by Links \etal \cite{Links-JPA35}, and Asorey \etal \cite{Asorey-et-al} are: 
\begin{eqnarray}
\frac{1}{g} &=&\sum_{i=1}^{L}\frac{{{\Omega }_{i}}}{2{{\varepsilon }_{i}}-{\
v_{\alpha }}}+\sum_{\beta \neq \alpha }^{M}\frac{2}{{v_{\alpha }}-{v_{\beta }
}}+\sum_{\gamma =1}^{M-T}\frac{1}{{w_{\gamma }}-{v_{\alpha }}}
\label{Links eqs.} \\
0 &=&\sum_{\alpha =1}^{M}\frac{1}{{v_{\alpha }}-{w_{\gamma }}}+\sum_{\delta
\neq \gamma }^{M-T}\frac{1}{{w_{\gamma }}-{w_{\delta }}}, \quad
E =\sum_{\alpha =1}^{M}{v_{\alpha }}. \nonumber
\end{eqnarray}
The spectral parameters $v_{\alpha }$ have the same meaning as pair energies. The $w_{\gamma }$ parameters are related to the iso-spin symmetry of the proton-neutron pairing. By drooping the terms that contain  the $w_{\gamma }$ parameters one arrives at the Richardson's exactly solvable pairing for one type of particles \cite{Richardson}. If one considers only one pair case ($M=1=p$) in $L$ single particle levels with double degeneracy of each single particle level $i$ ($\Omega_i=(2j+1)/2=1$)  then one has the
one pair energy eigenvalues of the EPI (\ref{H_extPM}) and the standard pairing (\ref{H_P}):
\begin{eqnarray*}
 E =z, ~ \frac{1}{g} &=&\sum_{i=1}^{L}\frac{{1}}{2{{\varepsilon }_{i}}-{z}}.
\end{eqnarray*}
This is a special case of the ($p=1$) solution for the extended pairing model \cite{Feng}:
\begin{equation}
E_p^{\zeta}=z^{\zeta}-G(p-1), ~~ 
\frac{1}{G}=\sum_{i_1 \ldots i_p}\frac{1}{E_{i_1\ldots i_p}- z^{\zeta}}, ~~ 
E_{i_1\ldots i_p}=\sum_{n=1}^p 2\varepsilon_{i_n}.
\label{E_extPM}
\end{equation}
In the above equations, we intentionally kept the notation for the 2-body pairing and the A-body pairing slightly different to emphasize their different structures.

\subsection{Binding Energy of the Sn and Pb Isotope Chains}

Deformation is common in very heavy nuclei and this often justifies the success and application of the Nilsson model. For the purpose of our model, we use deformation parameters from Ref. \cite{Moller&Nix} and experimental binding energies from Ref. \cite{AudiG}. Theoretical relative binding energies (RBE) are calculated relative to a specific core, $^{152}$Yb, $^{100}$Sn, and $^{208}$Pb for the cases considered. The RBE of the nucleus next to the core is used to determine an energy scale for the Nilsson single-particle energies. For an even number of neutrons, we considered only pairs of particles. For an odd number of neutrons, we apply Pauli blocking to the Fermi level of the last unpaired fermion and consider the remaining fermions as if they are an even fermion system.  The valence model-space consists of the neutron single-particle levels between two closed shells with magic numbers 50-82 and 82-126. By using (\ref{E_extPM}), values of $G$ are determined so that the experimental and theoretical RBE match exactly. 

\begin{figure}[htb]
\centering
\includegraphics[scale=0.46]{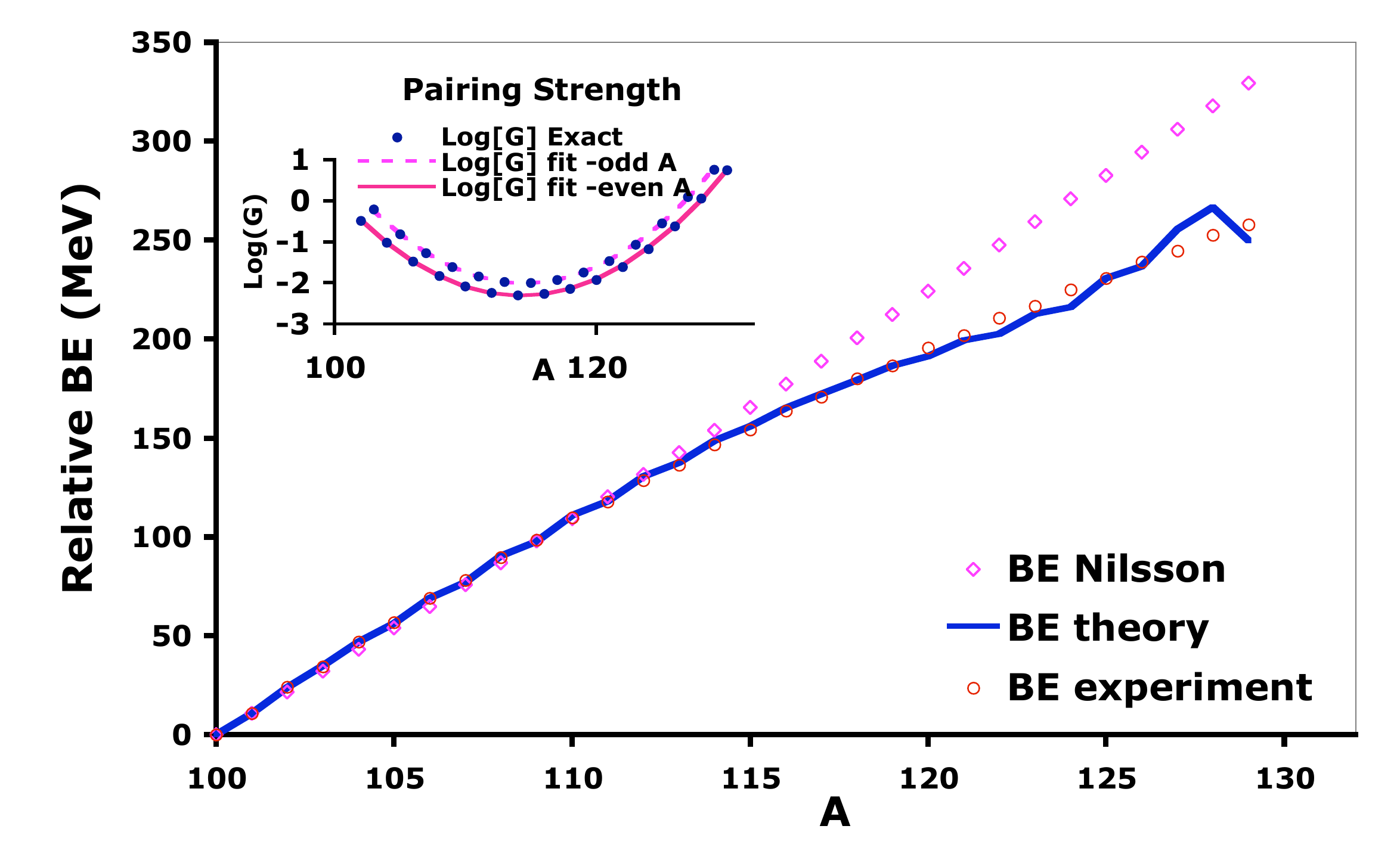}
\caption[]{Binding energies (BE) of the Sn isotopes relative to the BE of $^{100}$Sn core.}
\label{Sn_Isotopes_particles}
\end{figure}

Here we discuss mostly the Sn isotopes since the Pb and Yb isotopes were discussed in more details  in Ref. \cite{EPJ05} and Ref. \cite{VGG2004}.  
In Figure \ref{Sn_Isotopes_particles} are shown the results for Sn as calculated by using the $^{100}$Sn as core and zero RBE nucleus. The single-particle energy scale is set by the binding energy of $^{101}$Sn.  The inset shows the fit to values of $G$ that reproduces the experimental data exactly. The two fitting functions are: $\ln(G(A))=365.0584 - 6.4836 A + 0.0284A^2$ and $\ln(G(A))=398.2277 - 7.0349 A + 0.0307 A^2$ for even/odd values of $A$. The solid line gives the theoretical RBE of the Sn isotopes using these fitting functions.

\begin{figure}[htb]
\centering
\includegraphics[scale=0.48]{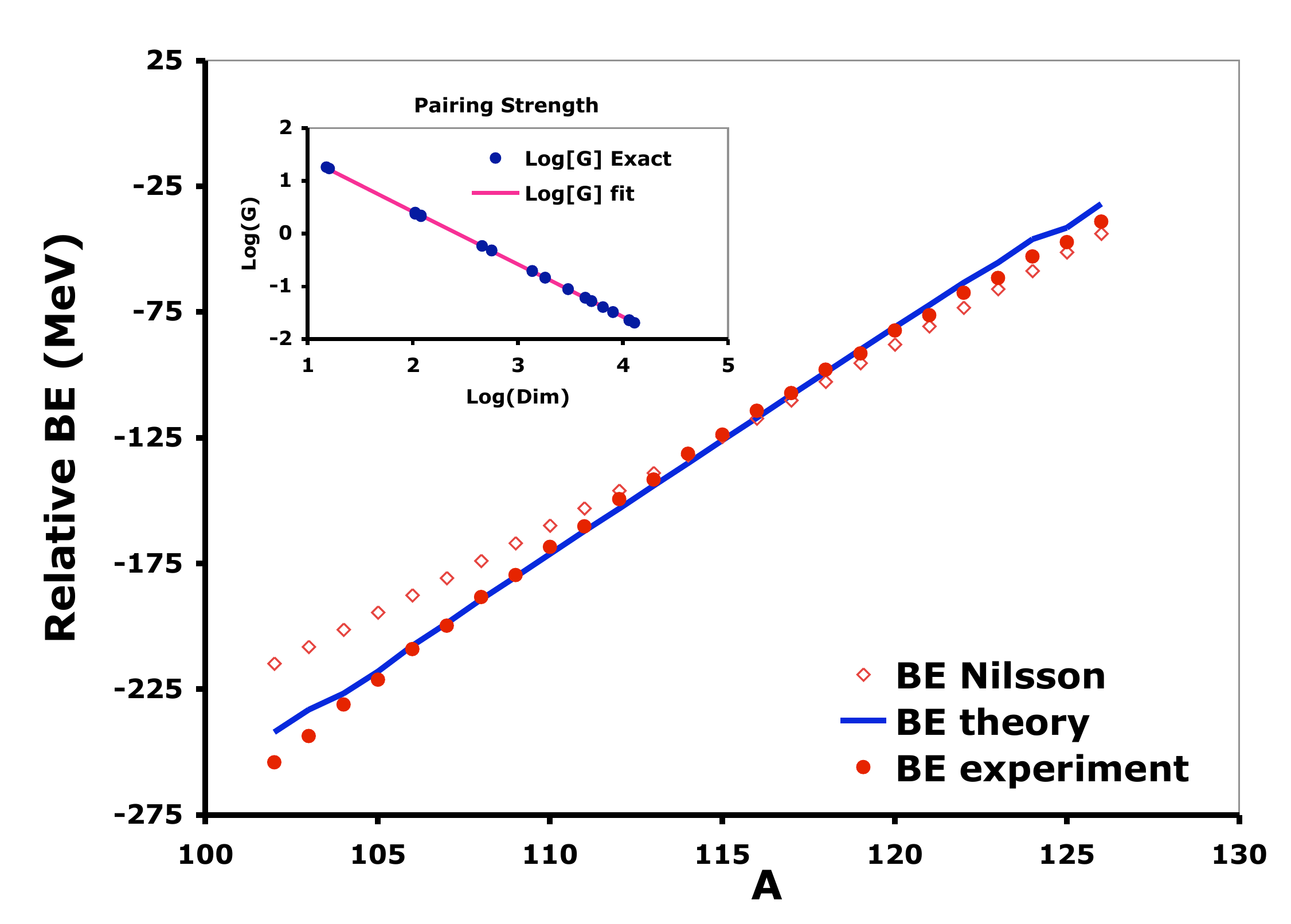}
\caption[]{Binding energy of the Sn isotopes relative to doubly magic $^{132}$Sn core.}
\label{Sn_Isotopes_holes}
\end{figure}

The Sn isotope chain is unique in the sense that  there are two doubly magic members, the $^{100}$Sn and $^{132}$Sn, that allows us to use $^{132}$Sn as zero RBE system as well. In Figure \ref{Sn_Isotopes_holes} are shown the results for Sn isotopes when using $^{132}$Sn as zero RBE system. In this case, there are again good even/odd quadratic dependence of the $\ln(G(A))$, however, as for Pb case \cite{EPJ05} there is a simpler expression that works for even and odd systems simultaneously. In this case we have $G(A)=\alpha\dim(A)^{-\beta}$ with $\alpha=259.436$ and $\beta=0.9985$.

\section{Conclusion}

In this paper we have presented evidence for the need to use accurate proton and neutron masses, or at least a properly weighted nucleon mass (\ref{Nmass_new}),  in order to improve on the accuracy of the binding energy of light nuclei as computed with the next generation computer codes. This will also allow better understanding of the NNN-, NNNN-, and A-body interactions in nuclei either derived from ChPT or from a phenomenological considerations. Therefore, one has to build A-body computational technology in the next generations of nuclear modeling codes.

While the motivation for considering A-body interaction in the light-nuclei is strong as based on the ChPT QCD derived interactions, one is left to wonder if A-body interactions are also relevant to heavy nuclei. The results obtained with the help of the Extended Pairing Interaction, in particular the Sn isotopes discussed here, seem to confirm the idea that A-body interactions are needed to understand better the binding energy of heavy nuclei. Often the imagination cannot capture all the possible implications and uses of an exactly solvable  model. Beside the current applications of the EPI, one can also see that it would be a useful verification tool for A-body computational codes as well. 

\section*{Acknowledgements}
V.G. Gueorguiev is grateful to his colleagues from the Bulgarian Academy of Sciences for the moral support and scientific encouragement, for their interest in his research, and for the many opportunities over the years to attend and present his research at their regular scientific meetings that they run very successfully over the years despite of the difficult economic times. J.P. Vary acknowledges partial support from DOE (grant DE-FC02-87ER-40371) and from NSF (grant NSF0904874). J.P. Draayer acknowledges partial support from the DOE (Grant DE-SC0005248), NSF (Grant NSF-0904874) and Southeastern Universities Research Association). This work was partly performed under the auspices of the U. S. Department of Energy by the University of California, Lawrence Livermore National Laboratory under Contract No. DE-AC52-07NA27344.

\end{document}